\documentclass{WileyMSP-template}
\usepackage[super,square,sort&compress]{natbib}
\usepackage[font={sf},labelfont=bf]{caption}
\usepackage{stmaryrd}
\usepackage{xcolor}
\usepackage{graphicx}
\usepackage{rotating}
\usepackage[colorlinks,citecolor=blue,linkcolor=blue]{hyperref}

\makeatletter
\DeclareRobustCommand\citenum
   {\begingroup
     \NAT@swatrue\let\NAT@ctype\z@\NAT@partrue\let\textsuperscript\NAT@spacechar
     \NAT@citexnum[][]}
\makeatother

\begin{document}
\baselineskip18pt

\title{\textbf{Compliant Lattice Modulations Enable Anomalous Elasticity in Ni-Mn-Ga Martensite}}

\maketitle

\bigskip


\author{Krist\'{y}na Rep\v{c}ek,}
\author{Pavla Stoklasov\'{a},}
\author{Tom\'{a}\v{s} Grabec,}
\author{Petr Sedl\'{a}k,}
\author{Juraj Olej\v{n}\'{a}k,}
\author{Mariia Vinogradova}
\author{Alexei Sozinov,}
\author{Petr Ve\v{r}t\'{a}t,}
\author{Ladislav Straka,}
\author{Oleg Heczko,}
\author{Hanu\v{s} Seiner*}

\bigskip
\begin{affiliations}
K. Rep\v{c}ek, P. Stoklasov\'{a}, T. Grabec, P. Sedl\'{a}k, H. Seiner*\\
Institute of Thermomechanics of the Czech Academy of Sciences, Prague 8, 18200, Czech Republic
\\{}*Email Address: hseiner@it.cas.cz\\

\medskip
J. Olej\v{n}\'{a}k\\
Faculty of Nuclear Sciences and Physical Engineering, Czech Technical University in Prague, Prague 2, 12000, Czech Republic\\

\medskip
M. Vinogradova, A. Sozinov\\
Material Physics Laboratory, Lappeenranta-Lahti University of Technology (LUT), Lappeenranta, 53850, Finland\\

\medskip
P. Ve\v{r}t\'{a}t, L. Straka, O. Heczko\\
FZU -- Institute of Physics of the Czech Academy of Sciences, Prague 8, 18200, Czech Republic\\

\end{affiliations}

\bigskip


\bigskip

\sloppy

\setlength\parindent{20pt}
\baselineskip15pt
\renewcommand{\arraystretch}{2}

\sloppy

\setlength\parindent{20pt}
\baselineskip15pt
\renewcommand{\arraystretch}{2}
{\bfseries Abstract --} High mobility of twin boundaries in modulated martensites of Ni-Mn-Ga-based ferromagnetic shape memory alloys holds a promise for unique magnetomechanical applications. This feature has not been fully understood so far, and in particular it has yet not been unveiled what makes the lattice mechanics of modulated Ni-Mn-Ga specifically different from other martensitic alloys. Here, results of dedicated laser-ultrasonic measurements on hierarchically twinned five-layer modulated (10\,M) crystals fill this gap. Using a combination of transient grating spectroscopy and laser-baser resonant ultrasound spectroscopy, it is confirmed that there is a shear elastic instability in the lattice, being  significantly stronger than in any other martensitic material and also than what the first-principles calculations for Ni-Mn-Ga predict. The experimental results reveal that the instability is directly related to the lattice modulations. A lattice-scale mechanism of dynamic faulting of the modulation sequence that explains this behavior is proposed; this mechanism can explain { the extraordinary mobility of twin boundaries in 10 M.} 

%
%
%
%
%
%

\section{Introduction}

Modulated martensites of Ni-Mn-Ga-based Heusler alloys\cite{Pons_Acta_2000,Lanska_JAP_2004,Takeuchi_NaturMat_2003} are so far the only known materials exhibiting the stunning phenomenon of twin boundary supermobility, that is, the ability to achieve large reversible strains under very small mechanical or magnetic loads via  motion of twin interfaces\cite{Sozinov_APL_2011, Straka_Acta_2011, Straka_JAP_2013, Seiner_PSS_2022,Heczko_MRS_2022}. This feature is maintained also at high speeds of twin boundary motion\cite{Saren_Scripta_2016,Saren_Scripta_2017, Musiienko_Scripta_2019, Dana_JalCom_2021} and down to cryogenic temperatures \cite{Heczko_APL_2013}, both making 
Ni-Mn-Ga martensites unrivaled candidates for  various applications\cite{Karaca_AdvFunMat_2009, Smith_Acta_2014, Yin_Sensors_2016, Heczko_MRS_2022, Chmielus_NatureMat_2009, Kohl_MSF_2010, Saren_Microfluidics_2018}. The supermobility itself is { generally understood as} a proof of an extraordinary  mechanical instability of the modulated crystal lattice, but the character and the origin of this instability have not yet been elucidated. This lack of understanding also hinders the search for materials with similar properties but with higher transformation and Curie temperatures than those of Ni-Mn-Ga, that is, temperatures more favorable for applications.

Despite the extensive research on Ni-Mn-Ga modulated crystals during past two decades, even their very basic mechanical properties are not exactly known, such as the elastic constants $c_{ij}$. These would be needed for any relevant continuum-level modelling of the applications\cite{Mousavi_SMS_2017, Fan_PhM_2022, Shi_IEEE_2019}, or for simulating the phenomenon of supermobility itself using topological models \cite{Faran_SMS_2016,Karki_Acta_2020,Shilo_Acta_2021}. Most importantly, however, the anisotropy in elasticity is strongly related to the mechanical stability of the lattice\cite{Nakanishi_PMS_1980}, the weakening of which is a precursor to either the martensitic transition or the twin reorientation. Thus, the determination of $c_{ij}$ would help significantly in understanding the unique mechanical behavior.

The main difficulties in obtaining the elastic constants experimentally originate from the presence of deeply hierarchical twinned microstructures in modulated Ni-Mn-Ga crystals \cite{Schwabe_AdvFunMater_2021,Seiner_JMPS_2014}. Due to the supermobility, these microstructures may rearrange under external loads below 1 MPa, which leads to mixing of elastic and inelastic responses in any mechanical test. Furthermore, both the modulated lattice itself and the hierarchical laminates have only a limited number of symmetry elements, which means that multiple independent elastic constants are needed to describe fully their behavior. Perhaps for these reasons the only available data on elastic constants of modulated Ni-Mn-Ga come from Dai et al.\cite{Dai_JAP_2004}, from measurements performed long time before the supermobility and the hierarchical character of microstructures have been discovered. In line with the knowledge at that time, Dai et al.\cite{Dai_JAP_2004} assumed a simple tetragonal symmetry of the material, performing, thus, their measurements on an unknown mixture of monoclinic variants with different orientations of the modulation vector. That is why their results cannot bring the desired insight into the instability of the modulated lattice.

\section{Results and Discussion}
\subsection{Determination of Elastic Constants}
\begin{figure}[!t]
 \centering
 \includegraphics[width=\textwidth]{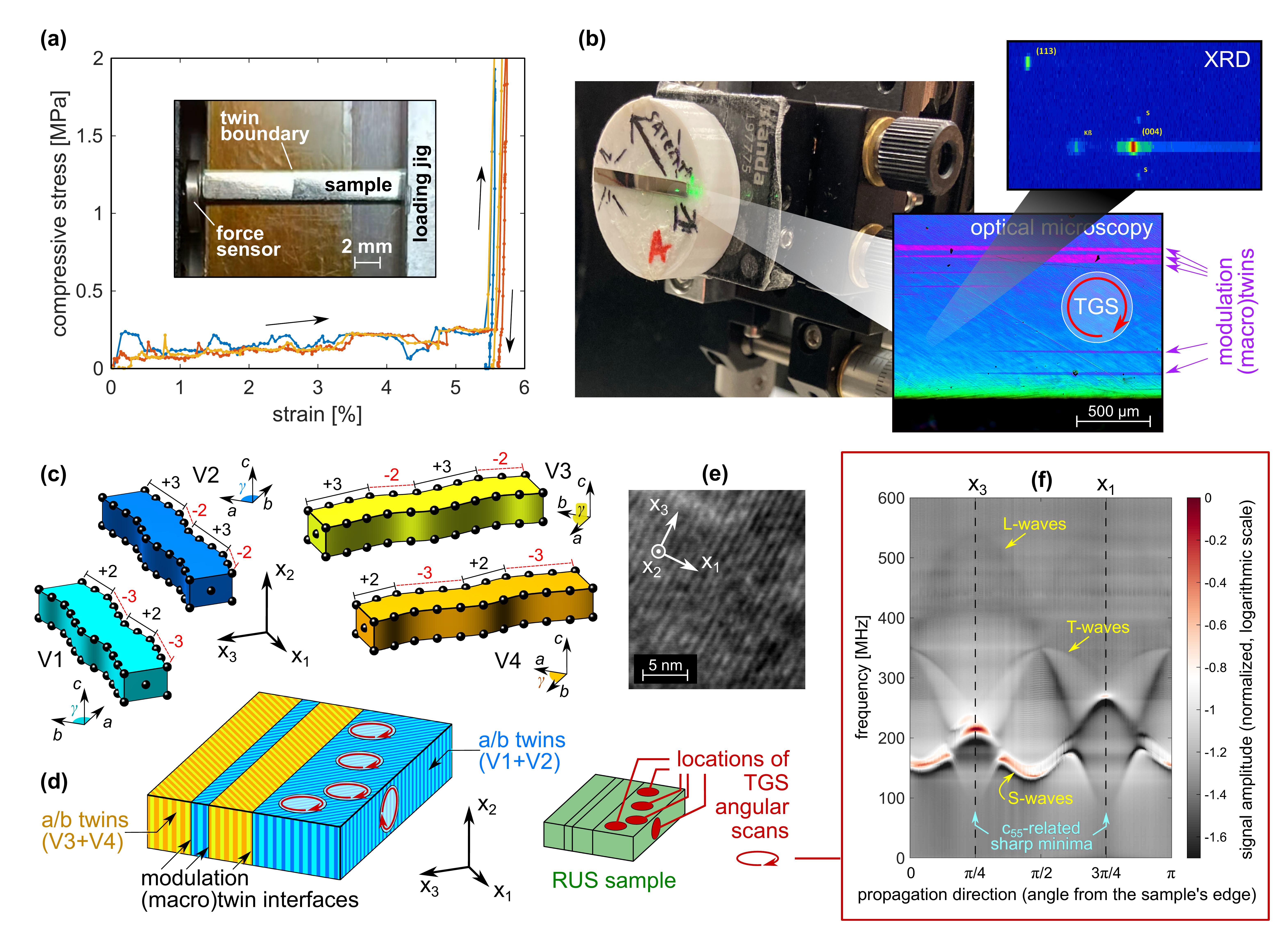}
 \caption{{{\bfseries (a)} twin supermobility in the examined material documented by repeated runs of compressive mechanical loading; the snapshot from the stress-strain experiment in the inset shows a single Type II interface propagating in the crystal;} { {\bfseries (c)} the crystal during the TGS measurement, with the green laser spot indicating the location and the size of the measured area; the locations were chosen not to intersect with the modulation (macro)twins based on optical microscopy observations (see the {\textcolor{blue}{Experimental section}}) and X-ray diffraction experiments (see the {\textcolor{blue}{Supplementary material}}).} {\bfseries (c)} four modulated 10\,M lattices, sharing the same orientation of the $c$-axis -- the variants V1 and V2 differ in the modulation sequence ($2\bar{3}2\bar{3}$ versus $3\bar{2}3\bar{2}$), the variants V3 and V4 differ from them by the direction of the modulation ($x_1$ versus $x_3$). The coordinate system is introduced based on V1, with $x_{1}$ being the modulation vector direction in V1 and $x_3$ being the direction of modulation polarization in V1; {\bfseries (d)} a schematic visualization of the arrangement of the individual variants in the examined material with respect to the used experimental characterizations; for simplicity the arrangement is shown on the geometry of the sample used for RUS; {{\bfseries (e)} the orientation of the chosen coordinate system $x_{i=1,2,3}$ introduced in (c,d) shown with respect to a TEM image of the 10 M modulated structure of the examined material}; {\bfseries (f)} an example of local TGS map for a 180$^\circ$ angular scan on a surface perpendicular to $x_2$. The labels L, T, and S denote longitudinal, transversal, and surface modes, respectively. Notice the clear mirror symmetry in frequency dispersions of all these modes with respect to the $x_1x_2$ and $x_2x_3$ planes, despite the measurements being performed on $a/b$ laminates that may break this symmetry  (see the {\textcolor{blue}{Supplementary material}} for a discussion).}
 \label{samplefig}
\end{figure}

For the experiments reported in this paper, we utilized laser-ultrasonics\cite{Scruby_LU_Textbook}, which means methods that generate and detect ultrasonic waves or ultrasonic vibrations in the crystal while not touching it other than by laser beams. By this advanced approach, we were able to probe locally the elastic behavior in different regions of the hierarchical microstructures, and to measure elastic constants without exposing the microstructures to stresses large enough to initiate the rearrangement. The experiments were done on a Mn-rich (Ni$_{50.2}$Mn$_{28.3}$Ga$_{21.5}$) five-layer modulated (10\,M) single crystal, a prototypical material with twin supermobility. { The used crystal exhibited the stress-strain behavior shown in Figure \ref{samplefig}(a), reaching strains of nearly 6\% under compressive stress varying between 0.1 and 0.2 MPa}. Using the procedure described in  the {\color{blue} Experimental section}, a defined hierarchical microstructure was created in the crystal prior to the measurements. { As confirmed by optical microscopy and X-ray diffraction (Figure \ref{samplefig}(b), see also the {\color{blue} Supplementary material}), this microstructure was} composed of four variants V1, V2, V3 and V4, defined in Figure \ref{samplefig}(c); these variants share the same orientation of the $c-$axis, but differ in the orientations of the $a-$ and $b-$axes and the modulation direction. The microstructure consisted of fine (micrometer-scale) laminates of $a/b$-twins spontaneously forming inside much broader (sub-millimeter-scale) modulation twins (Figure \ref{samplefig}(d), see Ref.\citenum{Straka_Acta_2011} for definitions of twinning types in Ni-Mn-Ga).

 Two different laser-ultrasonic methods were applied (see  the {\textcolor{blue}{Experimental section}}): \begin{enumerate} \item{}the elastic constants of $a/b$-laminates were locally probed at the surface by transient grating spectroscopy (TGS\cite{Stoklasova_ExpMech_2021}) using 10 $\mu$m wavelengths and the spot size of approximately 400 $\times$ 400 $\mu$m$^2$, the measurements were done in several different locations, covering both the surfaces oriented perpendicular to the $c-$axis of the material and those parallel to it (Figure \ref{samplefig}(d)); 
 \item{} the effective elastic constants of the whole twins-within-twins microstructure were determined by resonant ultrasound spectroscopy (RUS\cite{Sedlak_ExpMech_2014}), in which the wavelengths of the ultrasonic vibrations were comparable to the dimensions of the sample (approx. 3$\times$2$\times$1 mm$^3$). \end{enumerate}
 
 {X-ray diffraction was also used to determine the direction of the modulation vector (i.e. analyze the presence of (V1+V2) and/or (V3+V4) variants) in each location where the TGS angular scan was performed. It confirmed that there were no modulation twins in these locations, as indicated also by optical microscopy observations. In addition, the X-ray data confirmed that in each location the sample had the same commensurately modulated crystal structure. The details on the X-ray diffraction anaysis are provided in the {\color{blue}{Supplementary material}}}.

 \bigskip
Angular maps showing the dependence of frequency spectra of surface acoustic modes on the measurement direction were obtained from the TGS measurements (see Figure \ref{samplefig}(f) for an example). They always included weak peaks corresponding to surface-skimming longitudinal ultrasonic waves at frequencies above 400 MHz (marked as 'L-waves' in Figure \ref{samplefig}(f)). Then there were complex patterns of peaks and frequency bands corresponding to transverse waves ('T-waves', approximately between 200 and 350 MHz), and finally high-amplitude peaks representing surface  waves ('S-waves') appeared at around 150 to 200 MHz. In certain angular intervals, particularly slow S-wave modes were detected, decreasing in frequency often well bellow 100 MHz with simultaneously diminishing amplitudes when approaching propagation directions close to $x_3$ and/or $x_1$. These diminishing peaks represented waves travelling at velocities below 1 km.s$^{-1}$, which means much slower than for example acoustic waves in water. Such particularly slow modes indicated the presence of a shear instability in the lattice. 

\begin{figure}[!t]
\centering                                                                                                                                                                                                                                                                                                                                                                                                                                                                                                                                                                                                                                                                                                                                                      \includegraphics[width=\textwidth]{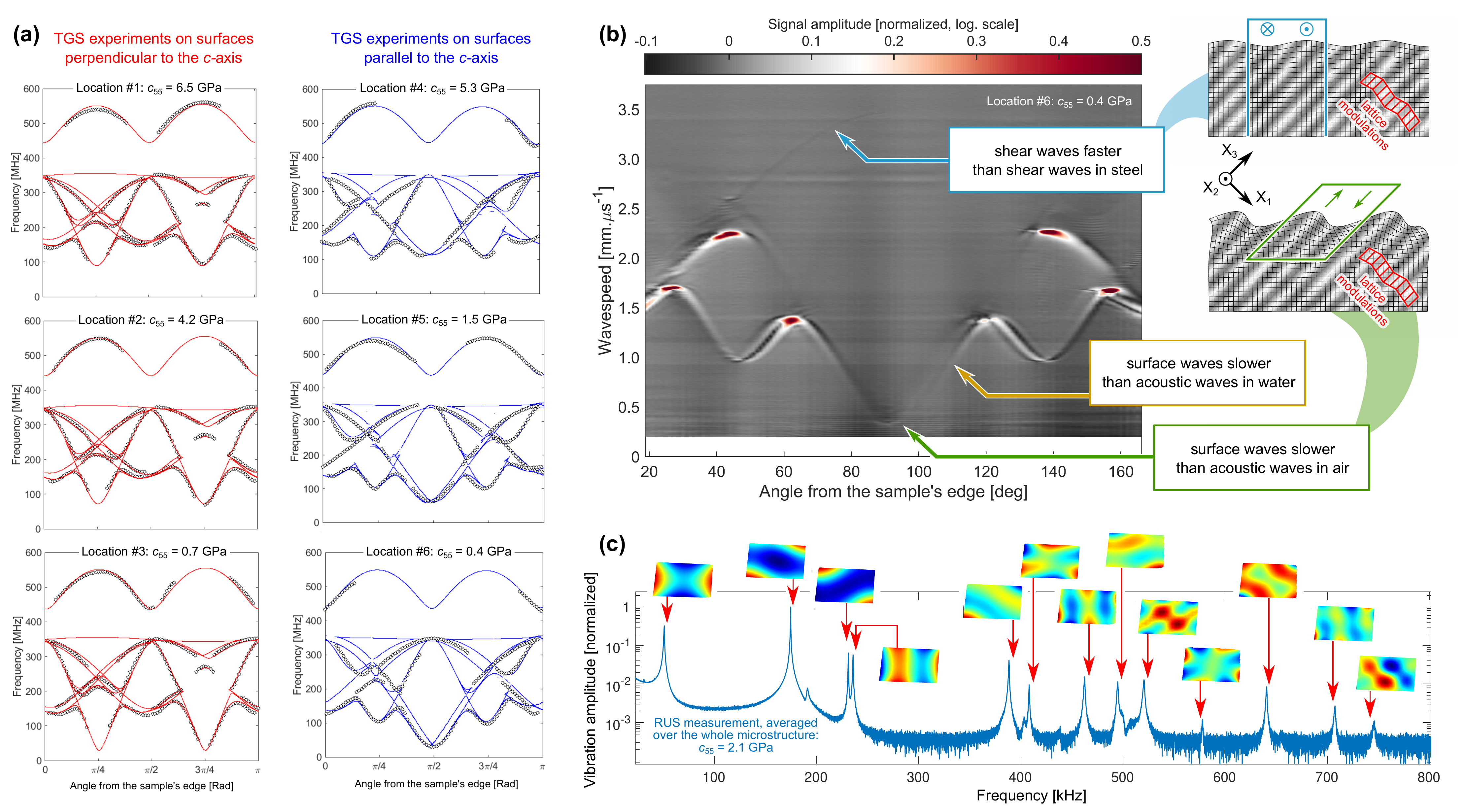}
\caption{Results of the laser-ultrasonic characterization. {\bfseries (a)} The resulting fit between experimental TGS map contours (circles) and calculated velocity maps for all considered modes (solid lines). The optimized sets of the elastic constants from location to location differed only in the value of $c_{55}$; all other elastic constants were shown to be the same over the whole sample. The experimental data were extracted from full 0$-$2$\pi$ angular maps, but are shown here gathered into the 0$-${}$\pi$ irreducible intervals. {\bfseries (b)} A selected angular interval of the TGS map showing the softest local behavior. For the vertical axis, the frequencies are recalculated into the corresponding speed of propagation of elastic waves. It is seen that for certain directions (angle $\approx$ 90$^\circ$), the S-wave mode reaches nearly 0.3 mm.$\mu$s$^{-1}$, while the fastest T-wave is close to 3.5 mm.$\mu$s$^{-1}$. On the right, these two extreme cases are visualized: The T-wave is a bulk wave polarized perpendicular to the modulations, detectable on the free surface only because a weak skimming effect\cite{Every_Ultras_2003}; the S-wave is polarized along the modulations and guided by the surface. {\bfseries (c)} RUS spectrum with all 13 detectable resonant modes. The red arrows mark frequencies for the given modes calculated theoretically for the best match ($c_{55}$); the modal shapes used for identification of individual modes were obtained using scanning laser vibrometry \cite{Sedlak_ExpMech_2014}.}
\label{S6}                                                                                                                                                                                                                                \end{figure}

TGS angular scans were measured at three different locations on the sample for each orientation of the free surface with respect to the $c-$axis. The contours of the individual modes (seen as peaks, anti-peaks or shoulders\cite{Scruby_LU_Textbook, Every_Ultras_2003}) extracted from these six TGS maps are plotted as dots in Figure \ref{S6}(a), and were used to calculate the elastic constants (see the {\textcolor{blue}{Experimental section}} for the used inverse numerical procedure). The unit cell of 10\,M Ni-Mn-Ga martensite is monoclinic, having thus 13 independent elastic constants $c_{ij}$\cite{Seiner_PSS_2022}. Since the measurements were performed on hierarchical microstructures, not all of these constants were experimentally available. As elaborated in detail in the {\textcolor{blue}{Supplementary material}}, the used experimental datasets enabled us to determine the orthorhombic part (9 elastic constants) of the tensor, listed in Table \ref{tab_elast}. Importantly, we were able to find a set of 8 elastic constants ($c_{11}$, $c_{12}$, $c_{13}$, $c_{22}$, $c_{23}$, $c_{33}$, $c_{44}$, and $c_{66}$) for which the calculated frequencies of L-waves and T-waves were matching all  experimental data with a good visual agreement (Figure \ref{S6}(a)), but to get the same agreement for the local minima for S-waves, significant variations of the constant $c_{55}$ between the locations were needed. The constant $c_{55}$ describes the shear stiffness of the lattice against shearing along the $x_2x_3$ plane and in the $x_3$ direction, that is, the same strains as those that create the lattice modulation itself. From the TGS results, we can conclude that this constant is extremely soft (down to 0.4 GPa, see Figure \ref{S6}(b) for the corresponding TGS map and for the visualization of the relation between the wave mode and the modulations), which manifests the instability of the lattice, and varies by an order of magnitude within one crystal. This means that the instability is locally more enhanced or suppressed; we discuss possible origins of such a heterogeneity at the end of this paper, and in more details in the {\textcolor{blue}{Supplementary material}}.  

The output of the RUS experiment was a vibrational spectrum of the whole sample containing the full hierarchical microstructure, i.e., $a/b$-twins within modulation macro-twins, Figure \ref{samplefig}(d). Because of a strong damping, only 13 vibrational peaks were detectable in the spectrum (Figure \ref{S6}(c), notice the logarithmic scale in the vertical axis, illustrating how rapidly the signal-to-noise ratio decreases with the increasing mode number). Due to the strong anisotropy, all these modes were possible to assume as dependent only on the softest shear constants in strongly anisotropic materials, and thus, they were suitable for determination of $c_{55}$ averaged over the whole sample volume, which resulted in $c_{55}=2.1$ GPa.  As expected, this value was between the maximum and minimum values of $c_{55}$ determined locally from the TGS measurements. For this value, the resonant frequencies of all detectable resonant modes matched very well the experimental peaks in the RUS spectrum, as seen in Figure \ref{S6}(c).

\begin{table}
 \caption{Elastic constants of 10~M Ni-Mn-Ga (orthorhombic part of the tensor) determined by laser-ultrasonic methods, and the corresponding literature data from first-principles calculations. The estimated error in the experimental data is $\pm 2$~GPa for TGS (for all constants except for $c_{55}$) and $\pm 0.2$~GPa for RUS.}
 \centering
 {
 \begin{tabular}{p{2.5cm}ccccccccc}
  \hline \hline
  {}&$c_{11}$&$c_{12}$&$c_{13}$&$c_{22}$&$c_{23}$&$c_{33}$&$c_{44}$&$c_{55}$&$c_{66}$\vspace{-.5cm}\\
  {}&[GPa]&[GPa]&[GPa]&[GPa]&[GPa]&[GPa]&[GPa]&[GPa]&[GPa]\\
  \hline
  {TGS}&246&153&59&155&117&250&95&0.4 -- 6.5$^{\rm a)}$ &102\\
  {RUS}& -- & -- & -- & -- & -- & -- & -- & 2.1 & -- \\ \hline
  {ab-initio\cite{Seiner_PSS_2022}}&238&139&59&187&111&262&89&18&88\\
  {ab-initio\cite{Ozdemir_JalCom_2010}}&271&135&56&194&127&265&91&15&100\\
  \hline \hline  
  \multicolumn{10}{l}{$^{\rm a)}$depending on the measurement location over the surface of the crystal}
 \end{tabular}}
\label{tab_elast}
\end{table}

\subsection{Discussion of the strong anisotropy}
\begin{figure}[!t]
\centering
 \includegraphics[width=0.9\textwidth]{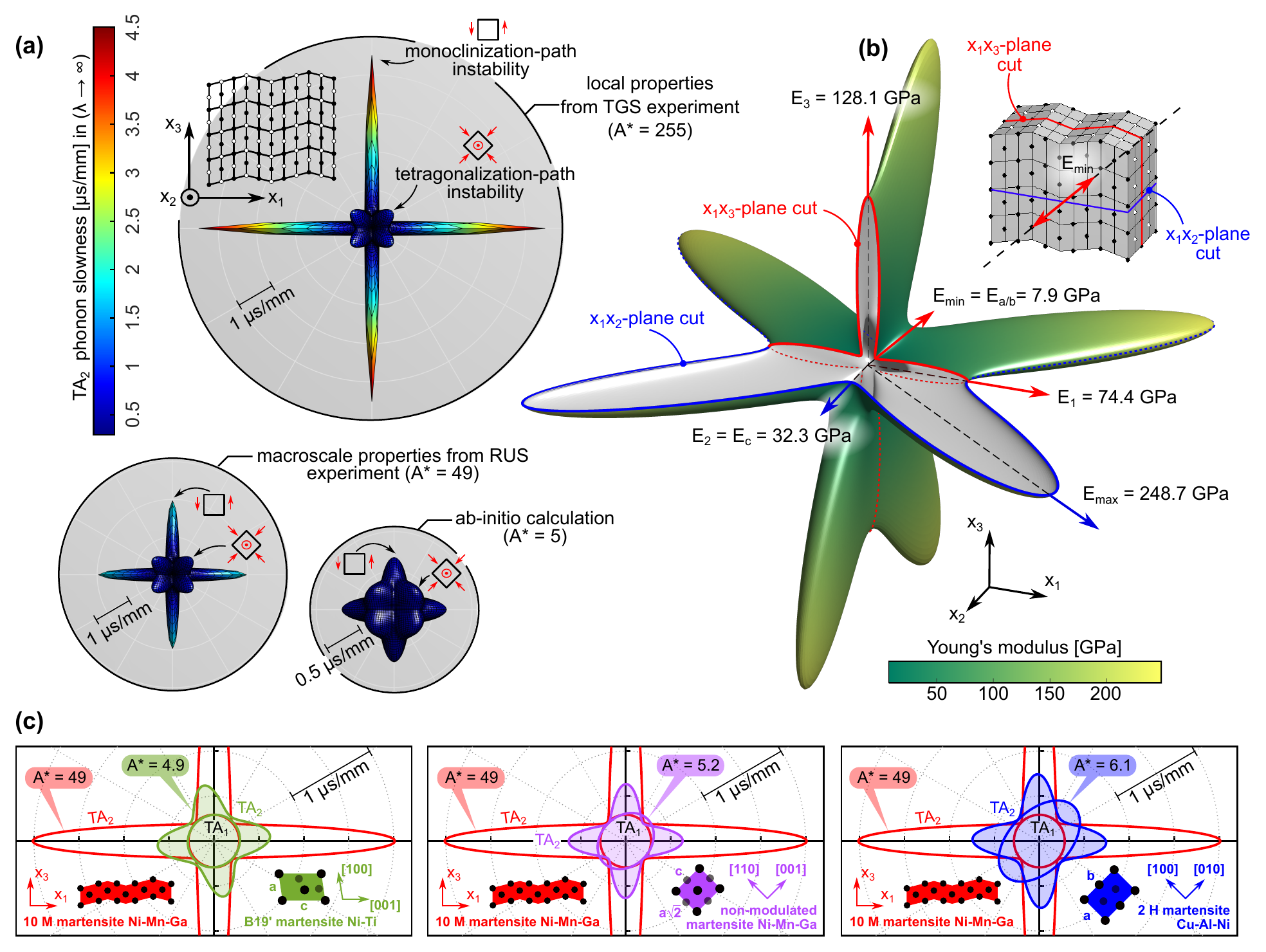}
 \caption{Extremely strong elastic anisotropy of 10~M Ni-Mn-Ga. {\bfseries (a)} Visualization of the elastic instability through three-dimensional plots of slowness (inverse value to phase velocity) of the softest (TA$_2$) acoustic phonon; $A^*$ stands for the generalized Zener anisotropy ratio. The last plot ($A^*$=5) shows how inaccurately is the instability captured by first-principles calculations, as discussed in Section 2.3. 
 {\bfseries (b)} Visualization of the strength of anisotropy (with $c_{55}$=2.1~GPa from RUS measurements, which means $A^*$=49) using a three-dimensional plot of Young's modulus with respect to the loading direction, and its cuts by principal planes. To make the cuts visible, one quadrant of the surface is removed. {{\bfseries (c)} Comparison of the strength of anisotropy of 10~M Ni-Mn-Ga and other prototypical martensite phases of shape memory alloys: Ni-Ti (monoclinic B19$^\prime$ martensite), tetragonal non-modulated martensite of Ni-Mn-Ga, and Cu-Al-Ni (orthorhombic 2H martensite), calculated using elastic constants from Refs. \citenum{Wagner}, \citenum{Sedlak_Scripta_2017}, and \citenum{Landa_APA_2009}, respectively. For all alloys, the cuts of the slowness surfaces of the faster (TA$_1$) and slower (TA$_2$) phonons in the $\lambda$$\rightarrow$$\infty$ limit are shown for crystallographic planes in which the strongest anisotropy is observed, and corresponding lattice orientations are depicted. It is seen that while the TA$_1$ slowness surface cut for 10 M Ni-Mn-Ga is comparable to those for all three other alloys, the strength of the instability indicated by the maxima on the  TA$_2$ slowness surface is significantly larger. This means that while the fastest shear waves in 10 M Ni-Mn-Ga travel at velocities very similar as in Ni-Ti, non-modulated Ni-Mn-Ga, or Cu-Al-Ni, the slowest shear waves in this material are three- to four-times slower than in other martensites.}}
 \label{anisotropy}
\end{figure}

The particularly small value of $c_{55}$ makes the 10~M hierarchically twinned crystal  one of the most anisotropic metallic materials ever reported (see Figure \ref{anisotropy}(a,b) for a visualization). For quantifying the anisotropy, the ratio $A^*$ between the stiffest and the softest shear modulus can be utilized (so-called generalized Zener anisotropy factor\cite{Lethbridge_Acta_2010}), reaching here $A^*=c_{66}/c_{55}=49$ on average, and $A^*=c_{66}/c_{55}=255$ in locations where $c_{55}$ falls to 0.4 GPa. In the extensive list of anisotropy ratios presented in \citenum{Lethbridge_Acta_2010}, the only higher reported value among metallic materials is $A^*\rightarrow2000$ for cubic InTl exactly at the transition temperature, where a complete collapse of the stability of the high-temperature cubic phase is observed. This is, however, not the case here, because 10~M martensite stays stable in the given alloy up to the \emph{austenite start temperature} $A_S\geq$ 323~K\cite{Straka_Acta_2011}, which means at least 30~K above the measurement temperature. For all martensite phases (i.e., low-temperature phases of ferroelastic alloys) either in Ref.\citenum{Lethbridge_Acta_2010} or in other literature, $A^*$ typically does not exceed 10 { (Figure \ref{anisotropy}(c))}. The value of $A^*>100$ that we observe locally by TGS is comparable to those of layered non-metallic materials such as graphite ($A^*=108$), which means that the modulation-related shearing is, in relative terms, as elastically soft as shearing of graphite layers along each other. 

The character and strength of the lattice instability is clearly seen from the slowness plot of the softest (TA$_2$) acoustic phonon in the long-wavelength limit, which means the slowest bulk elastic wavemode (Figure \ref{anisotropy}(a)). Sharp maxima along principal directions represent the extremely soft shears in the monoclinic plane, while the less sharp and much smaller maxima in direction inclined by $\pi/4$ from the monoclinic plane indicate instability with respect to the Bain tetragonalization path. This means that, contrary to expectation, the extreme shear instability is not related to the $c$$\leftrightarrow$$a$ reorientation path (i.e., the tetragonalization path), but to the monoclinization $b$$\leftrightarrow$$a$ path, which means it does not directly represent the twinning shear for supermobile $a/c$-interfaces. { The $c$$\leftrightarrow$$a$ tetragonalization path is related to the elastic constant $c^\prime$ representing the shears stiffness along the $c/a$ twinning planes. For $c_{55}=2.1$~GPa and other constants taken from Table \ref{tab_elast}, the value of this constant can be calculated as $c^\prime=10.2$ GPa, that is, significantly higher than the highest observed value of $c_{55}$.}

{ As shown by Saren et al.\cite{Saren_Scripta_2020} the mechanical stress needed for the $b$$\leftrightarrow$$a$ reorientation is also as low as 0.1 MPa. At the same time, the strain jump related to the $a/b$-twinning is by nearly two orders of magnitude smaller than for the $a/c$-twins, which means that the $a/b$-twins can move under a hundred times smaller driving force than the supermobile $a/c$-twins; this fully agrees with the observed extremely low value of $c_{55}$. However, a question arises, how the high mobility of $a/b$ twins and the low  $c_{55}$ relate to the supermobility phenomenon.}

The $a/c$-twins boundary motion is a complex microstructural process involving rearrangements of $a/b$ twins\cite{Seiner_JMPS_2014}, and thus, the observed softness in $c_{55}$ can significantly affect how energetically demanding the motion may be. As suggested in Ref.\citenum{Heczko_Acta_2013}, and discussed in relation to acoustic emission results in Ref.\citenum{Perevertov_PSS_2022}, the supermobile twin boundaries in 10~M Ni-Mn-Ga  do not exhibit any detectable pinning on lattice defects when they propagate through the crystal. It is plausible that the modulated lattice utilizes the energetically cheap $c_{55}$-related shears to overcome the pinning barriers, and the elastic constants of fine $a/b$ laminates reported in this paper are, therefore, an essential starting point for developing quantitative microstructural models of this process. { Even more importantly, however, the $c_{55}$-shears are related to the modulations at the nano-scale. At the end of this section, we propose a mechanism how the soft $c_{55}$ behavior can be achieved through energetically cheap rearrangements of the modulation sequence. This mechanism, at the same time, unlocks a new $c$$\leftrightarrow$$a$ reorientation path, that utilizes the $c_{55}$-related strains instead of those   $c^\prime$-related.}

\bigskip
In Figure \ref{anisotropy}(b), the strength of elastic anisotropy is visualized through a three-dimensional plot of Young's modulus, that is, the modulus representing the stiffness of the material with respect to unidirectional tension. The smallest value of the modulus is reached for tension along the $a$ or $b$ lattice directions ($E_{\rm a/b}$=7.9~GPa), for which the material can elongate utilizing the soft $c_{55}$-related shears most efficiently; $E_{\rm a/b}$ is more than 30 times smaller than the maximum value $E_{\rm max}$=248.7~GPa lying in the $x_1x_2$-plane cut. The values of $E_{\rm a/b}$ and $E_{\rm c}$ (in the $x_2$ direction) are very close to those reported from dynamical mechanical analysis (DMA) by Kustov et al.\cite{Kustov_Scripta_2020}. It is worth noting that the anisotropy of Young's modulus appearing spontaneously in the 10~M modulated Ni-Mn-Ga lattice is stronger than those of artificially built cellular materials and truss metamaterials, including those designed to exhibit strongly anisotropy \cite{Feng_MaterDes_2021,Khaleghi_MaterDesign_2021,Jiang_TWS_2023,Lan_IJMS_2023}.

\bigskip
Table \ref{tab_elast} provides an also important comparison between our experimental results and elastic moduli obtained by first-principles calculations reported in Refs. \citenum{Seiner_PSS_2022} and \citenum{Ozdemir_JalCom_2010}, where especially the former represents a state-of-the-art theoretical calculation, employing finely tuned $U-$Hubbard correction for electron localization.  It is observed that for all other constants than $c_{55}$ the experimental results fall within a $\pm{}20\%$ interval from the theoretical values, and also the character of the elastic anisotropy, that is, the ratios between individual constants, is in agreement between the calculation and the experiment. In contrast, $c_{55}$ differs by an order of magnitude, being $c_{55}\approx2$~GPa from our experiment and $c_{55}\approx20$~GPa from the calculation. As seen in the last plot in Figure \ref{anisotropy}(a), for this value of $c_{55}$ the maxima in the TA$_2$ slowness are much less pronounced, and the sharp difference between monoclinization an tetragonalization paths disappears. This means that the theoretical results not only give inaccurate predictions in terms of the strength of the elastic instability, but also preclude correct assessment of its character. On the other hand, the ab-initio calculations gave recently very accurate predictions for elastic constants of non-modulated Ni-Mn-Ga martensite \cite{Sedlak_Scripta_2017}, as well as for Ni-Mn-Ga-Co-Cu alloys exhibiting magnetically-induced reorientation\cite{Bodnarova_SMST_2021}. This means that the available first-principles models of the Ni-Mn-Ga-based Heusler structures have been correctly developed, and should be expected to give correct predictions also for the 10~M modulated structure.

The overestimation of the softest shear constant by ab-initio calculations, and thus, also the underestimation of the lattice instability of 10~M, is obvious also when comparing these theoretical results with previous available experimental data. Dai et al.\cite{Dai_JAP_2004} documented that even in a randomly twinned material (an unknown mixture of variants) the softest shear constant reached $c^*\approx5$~GPa, and, using X-ray diffraction measurements, Cejpek et al.\cite{Cejpek_Acta_2023} recently showed that under uniaxial tension, the spacing between specific planes in 10~M Ni-Mn-Ga martensite evolves as if the material has the initial Young's modulus as low as 0.7 GPa. Although these previous experimental results do not enable achieving any complete picture of the elastic anisotropy and lattice instability of 10~M, they confirm that there is a striking difference between the experiment and the theoretical prediction,  indicating that there exists a deformation mechanism in the 10~M lattice that is not captured by the currently available ab-initio tools, and which has not been so far considered in discussion of the supermobility phenomenon.

\begin{figure}[!t]
\centering
 \includegraphics[width=\textwidth]{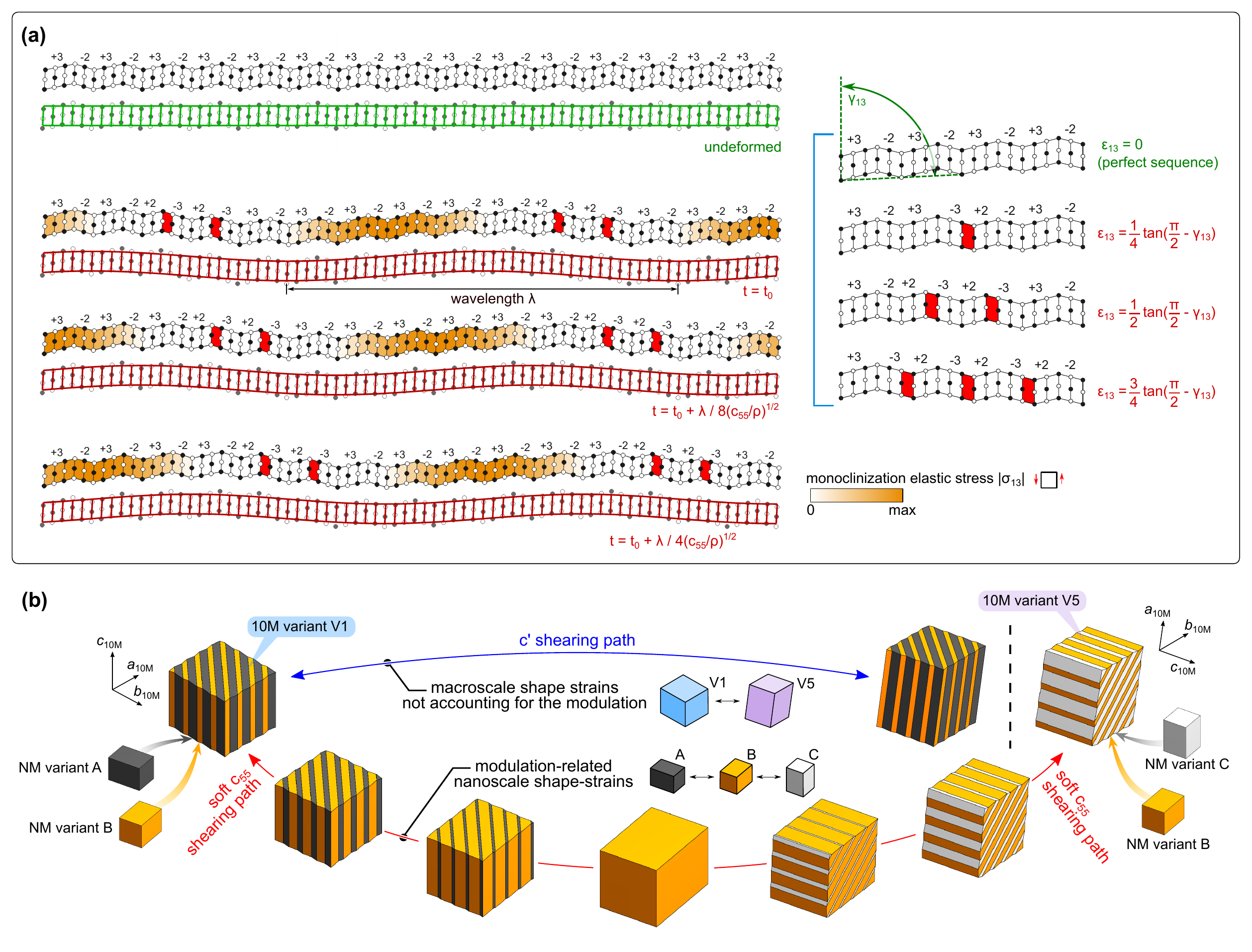}
 \caption{{\bfseries (a)} The proposed mechanism of dynamic creation and disappearance of the stacking faults in a planar shear wave traveling at velocity $v=\sqrt{c_{55}/\rho}$ in the $x_1$ direction. This mechanism can be understood as coupling between long-wavelength TA$_2$ phonons ($\lambda$ equals 20 interatomic spacing in the used visualization) and the short-wavelentgh phonons (the faults themselves, red-filled subcells). The left drawing shows the wave at three chosen time instants, with the solid red lines for each instant depicting the effective harmonic wave approximating the shear strains $\varepsilon_{13}$ in the lattice; on the right, examples of faulted sequences reducing the monoclinic distortion are shown. The faulting leads to effective inelastic monoclinization strains $\varepsilon_{13}$ that relax locally the elastic stresses  $\sigma_{13}$. The modulated lattice is visualized as variant V2 from Figure \ref{samplefig} with a $3\bar{2}3\bar{2}$ sequence, but in the used orthorhombic approximation V2 is not distinguishable from V1, with switched $a$ and $b$ orientations and a reversed $2\bar{3}2\bar{3}$ stacking sequence. {{\bfseries(b)}  Two possible $a$$\leftrightarrow$$c$ reorientation paths: the $c^\prime$-related path does not map the V1 lattice onto the V5 lattice because of the different orientations of the modulation; the $c_{55}$-related path utilizes the mechanism proposed in {\bfseries (a)} for a continuous connection between the lattices. Within the frame of the nano-twinning concept of modulations\cite{Kaufmann_NJP_2011}, the latter can be understood as a nanoscale reorientation between non-modulated (NM) variants A, B, and C, where the NM variant B is shared by the 10~M lattices V1 and V5.}}
 \label{elastic_faulting}
\end{figure}

We propose as an explanation that the ab-initio calculations do not assume any changes of the modulation sequence under external loads. Faults in the stacking can efficiently relax the energy of externally imposed $c_{55}$ shear strains, such as in a propagating shear wave sketched in Figure \ref{elastic_faulting}(a). Faulted sequences in 10~M NiMnGa martensite are known to be energetically cheap\cite{Kaufmann_NJP_2011}, and those reducing the effective monoclinic distortion may be even preferred over perfect sequences in terms of internal energy, because the modulation evolves from commensurate towards incommensurate with cooling\cite{Vertat_JPCM_2021} with a simultaneous reduction of the distortion. This leads us to the proposed mechanism in Figure \ref{elastic_faulting}(a), where the monoclinic distortion is quantified by the angle $\gamma_{\rm 10M}$. The monoclinization strains $\varepsilon_{13}$ carried by the wave induce elastic stresses $\sigma_{13}$ in regions where the  angle $\gamma_{\rm 10M}$ becomes more acute, but are relaxed by dynamic stacking faults in regions where $\gamma_{\rm 10M}\rightarrow{}\pi/2$. The relaxation mechanics softens the response, which results in the very small value of the determined $c_{55}$, and the corresponding very slow propagation of the shear wave in the given direction.

Also, one can assume that the creation of stacking faults is a dissipative process, which may explain why the uniaxial-stress vibrations along the direction of the minimal Young's modulus $E_{a/b}$ are much more damped than those along the $c-$axis, as reported in Ref.\citenum{Kustov_Scripta_2020}. Finally, the concept of inelastic straining through reversible stacking faults or other modifications of the modulation can also give a plausible explanation for the heterogeneity in $c_{55}$ observed from local TGS measurements. Pre-existing faults in the modulation (including $a/b$-twins as mirror reflections in the stacking sequence), other defects such as mosaicity or local chemical heterogeneities, or local internal stresses from polishing may diminish the ability of the 10~M lattice to utilize the faults for soft inelastic shearing, and in such locations, the response of the material becomes stiffer; we discuss these features further in the {\textcolor{blue}{Supplementary material}}.

\bigskip
{ The mechanism proposed in Figure \ref{elastic_faulting}(a) has been hypothesized based on the observed soft elastic behavior and other known features of the 10~M Ni-Mn-Ga martensite, and further dedicated experiments and theoretical calculations would be needed to prove its validity.} { An important contribution to this discussion will be the fact that the proposed mechanism enables explaining directly the relation between the supermobility and the $c_{55}$ elastic coefficient. This is outlined in Figure \ref{elastic_faulting}(b):
We consider two variants of 10~M martensite (V1 and V5) with different orientations of the $c-$axis; these two variants can form a supermobile Type II interface\cite{Straka_Acta_2011}. In the terms of strains, the variants V1 and V5 differ both in the macroscale shape strains (that is, the strains observed during the reorientation in Figure \ref{samplefig}(a)) and the nanoscale modulation strains, since the modulations in V1 and V5 run along different directions and in different planes. For simplicity, the modulations in Figure \ref{elastic_faulting}(b) are approximated using the nano-twinning concept\cite{Kaufmann_NJP_2011}, that is, assuming each modulated variant as a nano-laminate of two variants of non-modulated (NM) tetragonal martensite. Here, V1 is composed of NM variants A and B, while V5 is composed of B and C.

The macroscale strains are related to the elastic coefficient $c^\prime$, which has been shown above to be much stiffer than $c_{55}$, and are decisive for the orientation of the $a/c-$twinning plane through the compatibility conditions\cite{Straka_Acta_2011}. These conditions do not, however, fully determine the straining the lattice undergoes at the twinning plane, that is, in a twinning dislocation propagating along the interface\cite{Shilo_Acta_2021}. In particular, it is seen that the macroscale strains do not map the lattice of V1 directly onto the lattice of V5 (the $c^\prime$ shearing path in Figure \ref{elastic_faulting}(b)), because the modulation needs to be rebuilt.

On the other hand, if the soft elastic response is indeed caused by local faulting of the modulation sequence, an alternative pathway between V1 and V5 can be considered. The growth of the minor NM variant (B) in V1 at the expense of the major variant (A) is exactly the $c_{55}$-related shearing that reduces the monoclinic distortion, that is, the shearing relaxed by stacking faults in Figure \ref{elastic_faulting}(a). For symmetry reasons, the same mechanism can be considered for variant V5, where the minor NM variant (B) can grow at the expense of the major NM variant (C) \emph{via} the soft $c_{55}$ shears. As a result, a continuous pathway arises between V1 and V5 (soft $c_{55}$ shearing path in Figure \ref{elastic_faulting}(b)). This pathway represents rebuilding the modulation from one orientation into another, while the macroscale strains are achieved solely as a consequence of this rebuilding. As a result, the $c$$\leftrightarrow$$a$ reorientation may not be affected by the $c^\prime$ coefficient at all.} { It is plausible that this alternative pathway may be active also in other modulated martensite than 10~M, for example in seven-layer modulated 14~M martensite of Ni-Mn-Ga-Fe, where the supermobility of Type II twins has been recently reported\cite{Sozinov_Scripta_2020}.}

\section{Conclusions}

The experiments reported in the paper document the unique behavior of Ni-Mn-Ga 10~M modulated martensite in terms of its elastic properties. Despite being a metal, this material exhibits an extremely soft response with respect to shearing along specific crystallographic planes, resembling, thus, rather graphite or other Van der Waals materials. Such a behavior can be understood as { a consequence} of a strong instability of the lattice that makes the lattice easy to reorient into other variants of martensite, resulting in the observed twin boundary supermobility. { Contrary to expectation, the extremely soft elastic response is not associated with the macroscopic shape strains between the 10~M variants forming supermobile interfaces, but with nanoscale modulation-related strains.}  

Since the instability is not captured by first-principles calculations, it is plausible that it cannot be explained in terms of the stiffness of the interatomic bonds. In fact, the soft $c_{55}$-shearing is not anyhow related to the austenite-to-martensite transition path, and thus, there should not be any reason for its softening in the vicinity of the reverse transition temperature, which is something that would have been not hidden to the ab-initio approaches. Instead, we claim that the soft shearing is acquired through imperfections in the lattice modulations, which are known to appear spontaneously in Ni-Mn-Ga 10~M martensite. By faulting locally the stacking sequence, the lattice gains energetically cheap shear strains, which can efficiently relax the external loads. Identifying and understanding this mechanism may be the key finding for unravelling the twin supermobility itself. And, vice versa, it may help in the search for other materials than Ni-Mn-Ga exhibiting this stunning feature.

\section{Experimental section}
\subsection*{Material preparation and manipulation}
The material examined in this study was a single crystal of Ni-Mn-Ga produced by Adaptamat Ltd. (Finland). The chemical composition was determined as  Ni50.2Mn28.3Ga21.5 at.\% using X-ray fluorescence spectroscopy. The crystal was cut approximately along the $\{1\, 0\,0\}$
planes of the parent cubic austenite, having approximately 14$\times$2$\times$1 mm$^3$ in dimensions. {The  sample was ground mechanically in austenite  at 333 K up to P4000 grit, which was followed by  electropolishing at 258 K in an electrolyte consisting of a 3:1 solution of 60\% HNO$_3$ and ethanol, using 18 V power source and Pt-foil counter electrode.}
Prior to the measurements, all highly mobile twin interfaces (so-called $a/c$-twins) were driven out from the material using magnetic field, such that the resulting microstructure included only twins that do not alter the orientation of the magnetically easy $c-$axis, which are the so-called $a/b$-twins and modulation twins\cite{Straka_Acta_2011}.  After the field had been removed, the hierarchical microstructure in the sample was as is typical for 10\,M martensite\cite{Straka_Acta_2011,Seiner_JMPS_2014}, and as described in Figure 1. The magnetic field of 1.4 T was used to switch the $c-$axis orientation between states perpendicular and parallel to the electropolished surface, which enabled TGS measurements in both these states. { Since the sample was ground and electropolished in the austenite phase, it was possible to confirm the absence of the $a/c$-twins, as well as to observe the modulation (macro)twins using optical microscopy with differential interference contrast (DIC). As seen in Figure \ref{samplefig}(b), the small surface tilt between the V1+V2 and V3+V4 laminates enabled a clear visualization of the macrotwin bands, and allowed us to select the locations for TGS measurements, especially for the in-plane orientation of the $c-$axis. For the out-of-plane orientation of the $c-$axis, for which the surface tilt between V1+V2 and V3+V4 appeared only due to the slight misorientation of the polished surface from the principal plane, the DIC observation was not fully reliable, and the absence of the modulation (macro)twins in the regions chosen for TGS needed always to be confirmed from X-ray diffraction (XRD). XRD experiments were also used for both orientations of the $c-$axis to determine the orientation of the modulation direction in the selected location, as well as to confirm that the material had the same commensurate 10~M structure in each of the analyzed locations. A detailed description of the XRD characterization is provided in the {\textcolor{blue}{Supplementary material}}. }

After the TGS measurements, a smaller sample (approx. 3$\times$2$\times$1 mm$^3$) was cut from one end of the crystal and oriented using magnetic field such that the $c-$axis pointed perpendicular to the largest face of the sample. This sample was used for RUS experiments. { The remaining larger part of the crystal was then used for the stress-strain experiments shown in Figure \ref{samplefig}(a).}  

\subsection*{Laser-ultrasonic measurements}
All ultrasonic experiments were carried out at room temperature (295 K) and in zero magnetic field. TGS experiments were performed using the instrumentation described in detail in \citenum{Stoklasova_ExpMech_2021}, with a pulsed infrared laser for generation of transient standing-wave patterns and a continuous-wave green laser for the heterodyne detection. The used nominal wavelenght was $\lambda=10$ $\mu$m, the actual wavelength was calibrated on reference single-crystalline nickel. The spectra were recorded in frequency range 10 MHz -- 1 GHz, and with a 1$^\circ$ resolution in the angular scan.

For RUS experiment, the contact-less laser-based RUS arrangement described in detail in Ref.\citenum{Sedlak_ExpMech_2014} was utilized. This setup employs pulsed infrared laser for generation of vibrations, and scanning green-laser Doppler-type interferometer for detecting the vibrations and determination of modal shapes. The vibrational spectrum of the sample was recorded in a 50 kHz -- 2.5 MHz frequency range; however, due to strong damping, only few peaks at the lower end of the spectrum were distinguishable; that was, however, enough to determine the $c_{55}$ constant (see the {\textcolor{blue}{Supplementary material}} for a more detailed discussion). 

\subsection*{Inverse determination of the elastic constants}

For processing both the TGS and the RUS spectra in order to obtain the elastic constants, standard inverse-procedure approaches were utilized. This comprises first the identification of individual modes in the spectra, and then an iterative minimization of objective functions of form
\begin{equation}
 F(c_{ij})=\sum_{n=1}^{N}\left(\omega^{\rm exp.}_n-\omega^{\rm calc.}_n(c_{ij})\right)^2, \label{objective}
\end{equation}
where the superscripts ${\rm exp.}$ and ${\rm calc.}$ distinguish between experimentally determined frequencies and those calculated for iteratively improved guesses of the elastic constants $c_{ij}$. For TGS spectra, $\omega_{n}$ represented frequencies of different wave modes and in different directions of the angular scan. Typically, 2 to 5 peaks were identified for each propagation direction within the irreducible angular interval of 0 -- $\pi$, covering some combination of the S-, T- and L-modes. For RUS, $\omega_n$ represented different vibrational modes (first $N=11$ modes were detectable from the experiment).  
To obtain the corresponding sets $\omega^{\rm calc}_n(c_{ij})$, numerical Ritz-Rayleigh approaches were utilized for both TGS and RUS. Sensitivity analyses from Refs. \citenum{Stoklasova_ExpMech_2021} and \citenum{Sedlak_ExpMech_2014} were utilized to prove that the input experimental set carries enough information on the sought elastic constants. 

Because of the scatter in the constant $c_{55}$ discussed in the main text, we adopted the following strategy: 

Firstly, we noticed that all TGS maps obtained in different locations on the sample exhibited approximate mirror symmetries with respect to the $x_1x_2$, $x_1x_3$ and $x_2x_3$ planes, although the measurements were performed on $a/b$ laminates that violate these symmetries whenever the volume fraction ratio between V1 and V2 variants in the laminate differs from 50:50. These symmetries most probably result from the fact that in the given coordinate system the 10~M unit cell is nearly perfectly orthorhombic (because the monoclinic angle differs from 90$^\circ$ typically by less than 0.4$^\circ$ and the $a$ and $b$ lattice parameters differ by less than $0.5$\%), and thus, the modulation plane and the plane perpendicular to it remain approximately mirror planes of the microstructure, regardless of the volume fractions of the individual variants. This justifies the use of the orthorhombic part of the elastic tensor (from which all non-zero constants $c_{ij}$ are directly equal to the corresponding elastic constants of the monoclinic unit cell, as explained in the {\textcolor{blue}{Supplementary material}}). 

Secondly, we observed that the detectable T-waves and L-waves contours were nearly independent of the measurement locations, that is, they depended only on whether the experiments were performed on a surface perpendicular or parallel to the $c-$axis. However, the contours of the S-waves differed from location to location. Following this observation, we performed the inverse calculation of the elastic constants from the TGS results in two steps. \begin{enumerate}
                                                                                                                                                                                                                                                                                                                                                                                                                                                                                                                                                                                                                                                                                                                                                            \item L-waves and T-waves from all measurement points at both measured surfaces were taken as $\omega_n^{\rm exp.}$ to determine all orthorhombic elastic constants except for $c_{55}$. It was checked by a sensitivity analysis that L-waves and detectable T-waves are nearly independent of $c_{55}$. As a result, one set of all elastic constants other than $c_{55}$, that is  $\left(c_{11}, c_{12}, c_{13}, c_{22}, c_{23}, c_{33}, c_{44}, c_{66}\right)$, was obtained for the whole material.
                                                                                                                                                                                                                                                                                                                                                                                                                                                                                                                                                                                                                                                                                                                                                            
                                                                                                                                                                                                                                                                                                                                                                                                                                                                                                                                                                                                                                                                                                                                                            \item For each individual measurement location, the contours of L-, T- and S-waves were taken as $\omega_n^{\rm exp.}$ to determine full orthorhombic tensor of elastic constants, where only $c_{55}$ was sought by minimization of the objective function, while all other elastic constants were taken as fixed, using the result of the first step. 
                                                                                                                                                                                                                                                                                                                                                                                                                                                                                                                                                                                                                                                                                                                                                           \end{enumerate}

By using this approach, the elastic constants listed in Table \ref{tab_elast} were obtained,  with the resulting fit between measured and calculated frequencies seen in Figure \ref{S6}. It is seen that for all locations, a very good agreement is reached for all modes of waves. 

Finally, $c_{55}$ averaged over the volume of the RUS sample was sought by minimizing the misfit between 11 resonant frequencies and the calculated spectrum. An optimum match (less than 2\% difference for all involved frequencies) was reached for $c_{55}=$2.1 GPa. Two important points on this experiment are explained in the {\textcolor{blue}{Supplementary material}}: Firstly, the constant $c_{55}$ is invariant with both $a/b-$ and modulation twinning operations, and thus, the value determined on the hierarchical twinned microstructure can be directly related to single-variant elastic constants. Secondly, because of the exceptionally strong anisotropy, the lowest peaks in RUS spectra are nearly independent of all other elastic constants than $c_{55}$, that is, this constant can be determined reliably from the RUS measurement, even though other averaged constants of the hierarchical microstructure are not exactly known.

\section*{Acknowledgement}
{ The authors thank Dr. M. Vronka, FZU - Institute of Physics of the Czech Academy of Sciences, for TEM image used in Figure \ref{samplefig}.} This work has been financially supported by the Czech Science Foundation [project No. 24-10334S], by the Czech Technical University in Prague [project No. SGS22/183/OHK4/3T/14], and by the Operational Programme Johannes Amos Comenius of the Ministry of Education, Youth and Sport of the Czech Republic, within the frame of project Ferroic Multifunctionalities (FerrMion) [project No. CZ.02.01.01/00/22\_008/0004591], co-funded by the European Union. The authors from Finland acknowledge funding from the Academy of Finland [grant number 325910].

\section*{Data availability}
Data from laser-ultrasonic measurements and X-ray analyses are available from authors upon request.

%
%
%

\end{document}